\documentstyle[prd,aps,eqsecnum,preprint,tighten]{revtex}
\begin{document}
\preprint{\vbox to 50 pt
{\hbox{IHES/P/98/11}\hbox{CPT-98/P.3607}\vfil}}
\title{Light deflection by gravitational waves from localized sources}
\author{Thibault Damour}
\address{Institut des Hautes \'Etudes Scientifiques, F 91440
Bures-sur-Yvette, France\\
and DARC, CNRS Observatoire de Paris, F 92195 Meudon, France}
\author{Gilles Esposito-Far\`ese}
\address{Centre de Physique Th\'eorique\cite{UPR}, CNRS Luminy, Case
907, F 13288 Marseille Cedex 9, France}
\date{February 9, 1998}
\maketitle

\begin{abstract}
We study the deflection of light (and the redshift, or integrated time
delay) caused by the time-dependent gravitational field generated by a
localized material source lying close to the line of sight. Our
calculation explicitly takes into account the full, near-zone, plus
intermediate-zone, plus wave-zone, retarded gravitational field.
Contrary to several recent claims in the literature, we find that the
deflections due to both the wave-zone $1/r$ gravitational wave and the
intermediate-zone $1/r^2$ retarded fields vanish exactly. The leading
total {\it time-dependent} deflection caused by a localized material
source, such as a binary system, is proven to be given by the
quasi-static, near-zone quadrupolar piece of the gravitational field,
and therefore to fall off as the inverse cube of the impact parameter.
\end{abstract}
\draft

\section{Introduction}

The subject of light deflection, and/or light amplification, by
gravitational waves has a long but somewhat confusing history. Early
work \cite{zipoy,zipoybertotti,bertottitrevese}
correctly concluded to the absence of first-order effects increasing
linearly with the distance traversed by light within gravitational
waves. This result has recently been confirmed \cite{kaiserjaffe}, and
casts doubt on the possibility of detecting or constraining a
stochastic gravitational wave background by astronomical measurements
(be they astrometric or photometric). However, the works supporting
this pessimistic conclusion consider only purely transverse,
source-free, gravitational waves and often use an indirect formalism
(propagation equation for the local expansion rate of a light beam), so
that their impact on the problem of light deflection by gravitational
waves from localized sources is unclear. Recently, several authors have
considered the case where light rays pass close to a binary
gravitational-wave source and have suggested that, in such a
configuration, there could arise photometric \cite{labeyrie} or
astrometric \cite{fakir,durrer} effects proportional to the
gravitational-wave amplitude $h(b)$ evaluated at the impact parameter
$b$. If such effects $\propto h(b)$, decreasing only as $1/b$ in the
wave zone of the source, existed they might fall in the detectability
range of forthcoming optical interferometric arrays.

The purpose of the present paper is to study in detail the deflection
of light passing near any localized, non necessarily periodic,
gravitational-wave source (such as an inspiralling binary). We focus,
in particular, on the effect of the time-dependent gravitational field
associated with a varying quadrupole moment. As far as we know, our
treatment is the first one in the literature to work out the complete
effect of a time-dependent, retarded, quadrupolar field, $h(t,r)$,
which is given by a sum of terms having different fall-off properties
away from the source: $h(t,r) \sim a_1 (t-r) r^{-1} + a_2 (t-r) r^{-2}
+ a_3 (t-r) r^{-3}$. [To save writing we suppress here indices, though
our calculations take into account the full tensorial structure $h_{\mu
\nu} (t,{\bf x})$ of the gravitational field.] We find that the claims
of Refs.~\cite{labeyrie,fakir,durrer} are incorrect in
that the deflection is not of order $\alpha \sim h(b) \sim a_1 b^{-1} +
O (b^{-2})$, but falls off like $b^{-3} : \alpha \sim a_3 b^{-3}$. In
other words, the contributions to the deflection $\alpha$ of both the
purely wavelike field $\propto a_1 (t-r) / r$ and the faster falling
piece $a_2 (t-r) / r^2$ cancel out to leave only the contribution of
the (time-dependent) near-zone gravitational field $a_3 (t-r) / r^3$.
The resulting time-dependent deflection (which must be superposed on
the static effect of the total mass of the source) is much too small
(for reasonable impact parameters) to be of observational interest. The
same pessimistic conclusion applies to the other time-dependent effects
linked to $h(t,r)$: scintillation, variable redshifts and variable time
delays.

\section{Light deflection by a generic, time-dependent localized
gravitational source}

We work in the geometrical optics approximation. Let $\ell^{\mu} =
dz^{\mu} / d\xi$, $\mu = 0,1,2,3$, denote the tangent 4-vector to a
light ray $z^{\mu} (\xi)$ propagating in a curved spacetime $g_{\mu
\nu} (x^{\lambda})$ (with signature $-+++$). Here, $\xi$ denotes an
affine parameter along the light ray. The tangent vector $\ell^{\mu}$
is (by definition) ``light-like'' in the technical sense of
\begin{equation}
\ell^2 \equiv g_{\mu \nu} (z) \ell^{\mu} \ell^{\nu} = 0 \, ,
\label{eq2.1}
\end{equation}
and satisfies the geodesic equation $\ell^{\lambda} \nabla_{\lambda}
\ell^{\mu} = 0$, or, explicitly (with $\ell_{\mu} \equiv g_{\mu \nu}
\ell^{\nu}$)
\begin{equation}
\frac{d}{d\xi} \ell_{\mu} = \frac{1}{2} \ell^{\alpha} \ell^{\beta}
\partial_{\mu} g_{\alpha \beta} (z(\xi)) \, . \label{eq2.2}
\end{equation}
As the main aim of the present work is to clarify the deflecting effect
of locally generated bursts of gravitational waves on nearby passing
light rays, we shall formally consider that both the light source and
the observer are at infinity in a flat spacetime. In other words,
setting $g_{\mu \nu} (x) \equiv \eta_{\mu \nu} + h_{\mu \nu} (x)$, we
neglect the effects of $h_{\mu \nu}$ near the light source and near the
observer, and consider that the affine parameter $\xi$ varies between
$-\infty$ and $+\infty$. To first order in $h_{\mu \nu}$ the light
deflection is given by the 4-vector
\begin{equation}
\Delta \ell_{\mu} = \ell_{\mu} (+\infty) - \ell_{\mu} (-\infty) =
\frac{1}{2} \int_{-\infty}^{+\infty} d\xi \ell^{\alpha} \ell^{\beta}
\partial_{\mu} h_{\alpha \beta} (z_0^{\lambda} + \xi \ell^{\lambda})
\, . \label{eq2.3}
\end{equation}
In the right-hand side of Eq.~(\ref{eq2.3}) one can consider that
$\ell^{\alpha}$ denotes the constant, incoming light-like vector
$\ell^{\alpha} (-\infty)$, and we have replaced the photon trajectory
by its unperturbed approximation $z_{\rm unpert}^{\lambda} (\xi) =
z_0^{\lambda} + \xi \ell^{\lambda}$.

It will be technically very convenient to reexpress the deflection
(\ref{eq2.3}) in terms of the spacetime Fourier transform
$\widehat{h}_{\mu \nu} (k^{\lambda})$ of the gravitational field:
\begin{equation}
h_{\mu \nu} (x^{\lambda}) = \int \frac{d^4 k}{(2\pi)^4}
\widehat{h}_{\mu \nu} (k^{\lambda}) e^{ik\cdot x} \, . \label{eq2.4}
\end{equation}
Henceforth, we use the Minkowski metric $\eta_{\mu \nu}$ to raise and
lower indices, and make use of standard flat space notations, such as
$k \cdot x \equiv \eta_{\mu \nu} k^{\mu} x^{\nu} = {\bf k} \cdot {\bf
x} - \omega t$, $k^2 \equiv k \cdot k$, $h \equiv \eta^{\mu \nu} h_{\mu
\nu} , \ldots$ It is important to note that while $\ell^{\mu}$ (the
4-momentum of the impinging photon) is on-shell, $\ell^2 = \ell \cdot
\ell = 0$, the variable $k^{\mu}$ (4-momentum of the virtual gravitons
contributing to $h_{\mu \nu} (x)$) is generically off-shell, $k^2 \not=
0$.

After inserting Eq.~(\ref{eq2.4}) into Eq.~(\ref{eq2.3}), one can
perform the $\xi$-integration (using $\int d\xi \exp (ik \cdot z_0 + i
\xi k \cdot \ell) = 2\pi \delta (k \cdot \ell)$), with the result
\begin{equation}
\Delta \ell_{\mu} = i\pi \int \frac{d^4 k}{(2\pi)^4} k_{\mu}
\ell^{\alpha} \ell^{\beta} \widehat{h}_{\alpha \beta} (k^{\lambda})
e^{ik\cdot z_0} \delta (k \cdot \ell) \, . \label{eq2.5}
\end{equation}
This result is sufficient to show that {\it source-free} gravitational
wave packets do not deflect light. Indeed, any linearized, vacuum wave
packet $h_{\alpha \beta} (x)$ is a superposition of transverse, plane
waves propagating with the velocity of light. In other words, the
Fourier transform $\widehat{h}_{\alpha \beta} (k)$ of a source-free
wave packet contains a mass-shell delta function $\delta (k^2)$ and
satisfies (independently of the coordinate gauge) the transversality
condition
\begin{equation}
k^{\alpha} (\widehat{h}_{\alpha \beta} (k) - \frac{1}{2} \widehat{h}
\eta_{\alpha \beta}) = 0 \qquad (\hbox{on shell:} \ k^2 = 0) \, ,
\label{eq2.6}
\end{equation}
{}from which follows the consequence
\begin{equation}
k^{\alpha} k^{\beta} \widehat{h}_{\alpha \beta} (k) = 0 \qquad
(\hbox{on shell}) \, . \label{eq2.7}
\end{equation}
Coming back to Eq.~(\ref{eq2.5}), it is easy to see that when $k^{\mu}$
is on-shell, the delta function $\delta (k \cdot \ell)$, where {\it
both} $k^{\mu}$ and $\ell^{\mu}$ are on the light-cone, forces
$k^{\mu}$ to be parallel (or antiparallel) to $\ell^{\mu}$. The
deflection is then proportional to $\ell^{\alpha} \ell^{\beta}
\widehat{h}_{\alpha \beta} (k) \propto k^{\alpha} k^{\beta}
\widehat{h}_{\alpha \beta} (k)$ which vanishes because of
Eq.~(\ref{eq2.7}) (i.e. because of transversality). Therefore
\begin{equation}
\Delta \ell_{\mu} = 0 \, , \ \hbox{for any (localized) gravitational
wave packet}. \label{eq2.7n}
\end{equation}

This simple, algebraic argument makes it clear that the wave-like, $O
(1/r)$ part of the gravitational field generated by any local source
will give no contribution to the total deflection $\Delta \ell_{\mu}$
(when neglecting edge effects, and faster falling terms $O (1/r^2)$).
We need, however, to do more work to derive the explicit, non zero
value of $\Delta \ell_{\mu}$ generated by the contributions $O (1/r^2)
+ O (1/r^3) + \cdots$ to $h_{\mu \nu} (x)$. First, we need to relate
$h_{\mu \nu} (x)$ to the material source, i.e. to the (localized)
stress-energy tensor $T_{\mu \nu} (x)$. The linearized Einstein
equations, in harmonic gauge, read
\begin{equation}
\Box (h_{\mu \nu} - \frac{1}{2} h \eta_{\mu \nu}) = -16\pi G T_{\mu
\nu} \, , \label{eq2.8}
\end{equation}
or, in Fourier space,
\begin{equation}
k^2 \widehat{h}_{\mu \nu} (k) = + 16 \pi G (\widehat{T}_{\mu \nu} (k) -
\frac{1}{2} \eta_{\mu \nu} \widehat{T} (k)) \, , \label{eq2.9}
\end{equation}
where
\begin{equation}
\widehat{T}_{\mu \nu} (k) = \int d^4 x e^{-ik\cdot x} T_{\mu \nu} (x)
\, . \label{eq2.10}
\end{equation}
When dividing by $k^2$ to get $\widehat{h}_{\mu \nu} (k)$ from
Eq.~(\ref{eq2.9}), one needs to define carefully the singularity
structure at $k^2 = 0$, which is related to the boundary conditions
incorporated in the corresponding Green's function. The Fourier
transform of the usual {\it retarded} Green's function is $(k^2 - i
\epsilon k^0)^{-1}$, where $\epsilon$ is a positive infinitesimal, so
that
\begin{equation}
\widehat{h}_{\mu \nu} (k^{\lambda}) = 16 \pi G \frac{\widehat{T}_{\mu
\nu} (k) - \frac{1}{2} \eta_{\mu \nu} \widehat{T} (k) }{k^2 - i
\epsilon k^0} \, . \label{eq2.11}
\end{equation}
Note in passing that, in the decomposition ($P$ denoting the principal
part) \begin{equation} \frac{1}{k^2 - i \epsilon k^0} = P \frac{1}{k^2}
+ i \pi {\rm sign} (k^0) \delta (k^2) \, , \label{eq2.12}
\end{equation}
the second (on shell) term is the only one to contribute to the
``radiation'' Green function $G_{\rm retarded} - G_{\rm advanced}$
which defines a free wave packet associated to the source $T_{\mu \nu}$
and falling off at infinity like $1/r$. By the argument above we know
that this one-shell term will not contribute to $\Delta \ell_{\mu}$.
This shows that the deflection would be the same for the physical
retarded field $h_{\alpha \beta}^{\rm ret} (x)$, or for the acausal
fields $h_{\alpha \beta}^{\rm adv} (x)$ or $h_{\alpha \beta}^{\rm sym}
(x) = \frac{1}{2} (h_{\alpha \beta}^{\rm ret} (x) + h_{\alpha
\beta}^{\rm adv} (x))$. Let us continue working with the retarded field
(\ref{eq2.11}).

Inserting Eq.~(\ref{eq2.11}) into Eq.~(\ref{eq2.5}) we get (because of
the vanishing of $\ell^{\alpha} \ell^{\beta} \eta_{\alpha \beta}$)
\begin{equation}
\Delta \ell_{\mu} = 16 i \pi^2 G \int \frac{d^4 k}{(2\pi)^4}
\frac{k_{\mu} \ell^{\alpha} \ell^{\beta} \widehat{T}_{\alpha \beta}
(k)}{k^2 - i \epsilon k^0} e^{ik \cdot z_0} \delta (k \cdot \ell) \, .
\label{eq2.13}
\end{equation}

The energy-momentum conservation law $0=\nabla_{\nu} T^{\mu \nu} =
\partial_{\nu} T^{\mu \nu} + O (hT)$, or, in Fourier space and at
lowest order, $k^{\nu} \widehat{T}_{\mu \nu} (k) = 0$, gives explicitly
\begin{equation}
\widehat{T}_{0i} = - \frac{k^j}{k^0} \widehat{T}_{ij} \, ,
\label{eq2.14}
\end{equation}
\begin{equation}
\widehat{T}_{00} = - \frac{k^i}{k^0} \widehat{T}_{0i} = + \frac{k^i
k^j}{(k^0)^2} \widehat{T}_{ij} \, , \label{eq2.15}
\end{equation}
so that
\begin{equation}
\ell^{\alpha} \ell^{\beta} \widehat{T}_{\alpha \beta} (k) = (\ell^0)^2
\left( \frac{k^i}{k^0} - \frac{\ell^i}{\ell^0} \right) \left(
\frac{k^j}{k^0} - \frac{\ell^j}{\ell^0} \right) \widehat{T}_{ij} (k)
\, . \label{eq2.16}
\end{equation}
Let us henceforth split space and time and work in the center of mass
frame of the source (with the center of mass used as spatial origin).
The temporal origin is fixed by the requirement that the $\xi = 0$
event on the photon worldline is spatially closest to the center of
mass of the source and happens at coordinate time $t=0$. Technically,
this implies that $z_0^{\lambda} = (0, {\bf b})$ where the (vectorial)
impact parameter ${\bf b} \equiv b^i$ is orthogonal to the photon
3-momentum $\mbox{\boldmath$\ell$}\equiv \ell^i$. We can then
introduce usual polar coordinates $(\theta , \varphi)$ to parametrize
the direction of the 3-vector ${\bf k}$ with respect to a spatial
triad with $x$-axis along
${\bf b}$ and $z$-axis along $\mbox{\boldmath$\ell$}$, i.e.
$$
k^{\mu} = (\omega , k^1 , k^2 , k^3) = (\omega , k \sin \theta \cos
\varphi , k \sin \theta \sin \varphi , k \cos \theta) \, .
$$
Henceforth, $k,\ell ,\ldots$ denote the spatial lengths of the
3-vectors ${\bf k} , \mbox{\boldmath$\ell$} , \ldots$ and no longer
4-vectors as above. We denote also $\omega \equiv k^0$ (while $\ell^0
= \vert \mbox{\boldmath$\ell$} \vert = \ell$). From $k^{\mu}
\ell_{\mu} = {\bf k} \cdot \mbox{\boldmath$\ell$} -
\omega \ell^0 = k \ell \cos \theta - \omega \ell$ the delta function
in Eq.~(\ref{eq2.13}) reads
\begin{equation}
\delta (k^{\mu} \ell_{\mu}) = \frac{1}{k\ell} \delta \left( \cos \theta
- \frac{\omega}{k} \right) \, . \label{eq2.17}
\end{equation}
This implies that the integration on ${\bf k}$ is restricted to values
such that ${\bf k}^2 \geq \omega^2$. When writing Eq.~(\ref{eq2.13})
explicitly, we find it convenient to replace the $k$-integration in
$d^4 k = d \omega k^2 d k d (\cos \theta) d \varphi$ by an integration
over
\begin{equation}
u \equiv \sqrt{{\bf k}^2 - \omega^2} = k \sin \theta \, .
\label{eq2.18}
\end{equation}
Finally, inserting Eqs.~(\ref{eq2.16}) and (\ref{eq2.17}) into
Eq.~(\ref{eq2.13}) we get
\begin{eqnarray}
\alpha_{\mu} \equiv \frac{\Delta \ell_{\mu}}{\ell} & = & i
\frac{G}{\pi^2} \int_{-\infty}^{+\infty} \omega^{-2} d \omega
\int_0^{+\infty} u d u \int_0^{2\pi} d \varphi K_{\mu} (\omega , u ,
\varphi) e^{ibu\cos \varphi} \nonumber \\
&& [\cos^2 \varphi \widehat{T}_{11} (\omega , {\bf k}) + \sin^2 \varphi
\widehat{T}_{22} (\omega , {\bf k}) + 2 \sin \varphi \cos \varphi
\widehat{T}_{12} (\omega , {\bf k})] \, , \label{eq2.19}
\end{eqnarray}
where the denominator $({\bf k}^2 - \omega^2 - i \epsilon \omega)^{-1}$
cancelled with a contribution $\propto {\bf k}^2 - \omega^2$ in the
numerator (confirming the irrelevance of $\epsilon$, i.e. of the
Green's function boundary conditions), and where $K_{\mu}$ denotes the
value of $k_{\mu}$ when restricted by the delta function
(\ref{eq2.17}), namely
\begin{equation}
K_{\mu} (\omega , u , \varphi) = (K_0 , K_1 , K_2 , K_3) = (-\omega , u
\cos \varphi , u \sin \varphi , \omega) \, . \label{eq2.20}
\end{equation}
Note that the result (\ref{eq2.19}) is entirely expressed in terms of
the components of $\widehat{T}_{ij}$ pertaining to the $x-y$-plane,
i.e. the plane orthogonal to the direction of propagation of the
incoming light. This is again an aspect of the transversality of the
gravitational field.

\section{Light deflection by a time-dependent, quadrupolar
gravitational field}

To see better the physical content of the result (\ref{eq2.19}), let us
make the further approximation that the source internal motions are
nonrelativistic so that the time-dependent part of the external field
is well described by the quadrupolar approximation. Explicitly, this
means (in $x$-space), a field which reads (in a suitable harmonic
gauge, and after subtraction of the Schwarzschild-like, monopolar
piece)
\begin{eqnarray}
\widetilde{h}^{00} & = & + 2 G \partial_{ij} \left( \frac{D_{ij}
(t-r)}{r} \right) \, , \nonumber \\
\widetilde{h}^{0i} & = & - 2 G \partial_j \left( \frac{{\dot D}_{ij}
(t-r)}{r} \right) \, , \nonumber \\
\widetilde{h}^{ij} & = & + 2 G \frac{{\ddot D}_{ij} (t-r)}{r} \, .
\label{eq3.21}
\end{eqnarray}
Here $\widetilde{h}^{\mu \nu} \equiv h^{\mu \nu} - \frac{1}{2} h
\eta^{\mu \nu}$ satisfies $\partial_{\nu} \widetilde{h}^{\mu \nu} = 0$,
and
\begin{equation}
D_{ij} (t) = \int d^3 x x^i x^j T^{00} (t, {\bf x}) \label{eq3.22}
\end{equation}
is the quadrupole moment (with its trace). Note that, when expanding
the action on $r$ of the spatial derivatives in Eqs.~(\ref{eq3.21}),
the resulting retarded gravitational field contains a sum of
contributions of the form $a_1 (t-r) / r + a_2 (t-r) / r^2 + a_3 (t-r)
/ r^3$. The $1/r$ piece is the usual quadrupolar wave, the $1/r^3$
piece is a retarded version of the near-zone quadrupolar field $\left(
\widetilde{h}_3^{00} = 2 h_3^{00} = 2 G D_{ij} (t-r)
\partial_{ij} 1/r \right)$ and the $1/r^2$ piece plays a role in the
region intermediate between the near zone and the wave zone. Our
present calculation (done below in Fourier space) takes into account
all these contributions and allows one to study carefully the fall-off
properties of the light-deflection $\Delta \ell_{\mu}$ as a function of
the impact parameter.

In Fourier space, the quadrupolar approximation is easily seen (e.g. by
Fourier-transforming $\Box \widetilde{h}^{ij} = - 8 \pi G {\ddot
D}_{ij} (t) \delta^3 ({\bf x})$) to correspond to making the
approximation that
\begin{equation}
\widehat{T}_{ij} (\omega , {\bf k}) = \int d^4 x e^{-i{\bf k} \cdot
{\bf x} + i \omega t} T_{ij} (t,{\bf x}) \label{eq3.23}
\end{equation}
is independent of ${\bf k}$ \cite{weinberg}, so that (using the
standard virial theorem $\int d^3 x T_{ij} (t,{\bf x}) = \frac{1}{2}
\partial_t^2 \int d^3 x x^i x^j T_{00} (t,{\bf x})$ following from
$\partial_{\nu} T^{\mu \nu} = 0$)
\begin{equation}
\widehat{T}_{ij} (\omega , {\bf k}) \simeq \widehat{T}_{ij} (\omega ,
{\bf 0}) = \int dt e^{i\omega t} \int d^3 x T_{ij} (t,{\bf x}) =
-\frac{\omega^2}{2} D_{ij} (\omega) \, , \label{eq3.24}
\end{equation}
where $D_{ij} (\omega) \equiv \int dt e^{i\omega t} D_{ij} (t)$.

Under this approximation, one can explicitly perform the integrations
in our general result (\ref{eq2.19}). Indeed, all the $u$-integrals in
(\ref{eq2.19}) are of the form
\begin{equation}
U_n = \int_0^{\infty} du u^n e^{i(b\cos \varphi) u} = n! \left(
\frac{-1}{i b \cos \varphi - \epsilon} \right)^{n+1} \, ,
\label{eq3.25}
\end{equation}
with $n=1$ or $2$. Here, the positive infinitesimal $\epsilon$
(which is unrelated to the one entering the retarded Green's function)
is mathematically justified by appealing to distribution theory, or
physically justified by remembering that, in reality, $\widehat{T}_{ij}
(\omega , {\bf k})$, Eq.~(\ref{eq3.23}), must fall off to zero as
$\vert {\bf k} \vert \rightarrow \infty$, i.e. $u \rightarrow \infty$.
Using the result (\ref{eq3.25}), the $\varphi$-integrals in
Eq.~(\ref{eq2.19}) are of the form
\begin{equation}
\Phi_n = \int_0^{2\pi} d\varphi \left( \frac{\sin \varphi}{\cos \varphi
+ i \epsilon} \right)^n \, , \label{eq3.26}
\end{equation}
with $n=0,1,2$ or $3$. Clearly $\Phi_0 = 2\pi$, while $\Phi_1 = \Phi_3
= 0$ by symmetry. It remains only to evaluate $\Phi_2$ for which we
find
\begin{equation}
\Phi_2 = -2\pi + \frac{2\pi \epsilon}{\sqrt{1+\epsilon^2}} \, ,
\label{eq3.27}
\end{equation}
which tends to $-2\pi$ as $\epsilon \rightarrow 0$.

Finally, the two deflection angles $\alpha_1 = \Delta \ell_1 / \ell$,
$\alpha_2 = \Delta \ell_2 / \ell$, in the plane orthogonal to the
light ray (remember that the first axis is along ${\bf b}$, and the
second is parallel to $\mbox{\boldmath$\ell$} \times {\bf b}$), are
given by
\begin{equation}
\alpha_1 = - \frac{2G}{\pi b^3} \int_{-\infty}^{+\infty} d\omega
[D_{11} (\omega) - D_{22} (\omega)] = -\frac{4G}{b^3} [D_{11} (t_0) -
D_{22} (t_0)] \, , \label{eq3.28}
\end{equation}
\begin{equation}
\alpha_2 = + \frac{4G}{\pi b^3} \int_{-\infty}^{+\infty} d\omega D_{12}
(\omega) = + \frac{8G}{b^3} D_{12} (t_0) \, . \label{eq3.29}
\end{equation}
Here, $t_0$ ($=0$ in our coordinate system) denotes the date when the
light ray passes nearest to the source. The longitudinal fractional
change of the photon momentum $\alpha_3 = \Delta \ell_3 / \ell$ is
equal to the fractional change in photon energy $\alpha^0 = \Delta
\ell^0 / \ell = - \Delta \ell_0 / \ell$ and is given by
\begin{equation}
\alpha^0 = \alpha_3 = + \frac{iG}{\pi b^2} \int_{-\infty}^{+\infty}
d\omega \omega [D_{11} (\omega) - D_{22} (\omega)] = -\frac{2G}{b^2}
\frac{\partial}{\partial t_0} (D_{11} (t_0) - D_{22} (t_0)) \, .
\label{eq3.30}
\end{equation}

Let us note in passing that these results, here expressed in terms of
the quadrupole moment (\ref{eq3.22}) with its trace, depend only on the
trace-free quadrupole moment $Q_{ij} \equiv D_{ij} - \frac{1}{3}
D_{ss} \delta_{ij}$. This was a priori expected as it is well-known
that, modulo a coordinate transformation, the time-dependent
gravitational field external to any source depends only on $Q_{ij}
(t)$. We could everywhere replace $D_{ij}$ by $Q_{ij}$ but we will not
bother to do so.

The results (\ref{eq3.28})--(\ref{eq3.30}) can be encoded in a scalar
potential $V(z_0^{\lambda})$ which is essentially the gravitational
perturbation of the time delay between the light source and the
observer. Indeed, if we define
\begin{equation}
V (z_0^{\lambda}) = \frac{1}{2\ell} \int_{-\infty}^{+\infty} d\xi
\ell^{\alpha} \ell^{\beta} h_{\alpha \beta} (z_0^{\lambda} + \xi
\ell^{\lambda}) \, , \label{eq3.31}
\end{equation}
we see that Eq.~(\ref{eq2.3}) yields
\begin{equation}
\alpha_{\mu} \equiv
\frac{\Delta \ell_{\mu}}{\ell} = \frac{\partial}{\partial z_0^{\mu}} V
(z_0^{\lambda}) \, . \label{eq3.32}
\end{equation}
Using the integrals given above, it is easy to obtain
\begin{equation}
V (z_0^{\lambda}) = \frac{G}{\pi b^2} \int d\omega [D_{11} (\omega) -
D_{22} (\omega)] = \frac{2G}{b^2} [D_{11} (t_0) - D_{22} (t_0)] \, .
\label{eq3.33}
\end{equation}
Then, to compute Eq.~(\ref{eq3.32}) one needs to express $b,t_0$, as
well as the tensor projection $D_{11} - D_{22}$, as explicit functions
of $z_0^{\lambda} = (z_0^0 , z_0^i)$. This is achieved as follows: Let
the system's center of mass (c.m.) worldline be denoted $y^{\mu} (\tau)
= y_0^{\mu} + \tau u^{\mu}$, where $\tau$ is the c.m. proper time. In
spacetime, the impact parameter is a 4-vector $b^{\mu}$ which connects
$y^{\mu} (\tau)$ to the photon worldline $z^{\mu} (\xi) = z_0^{\mu} +
\xi \ell^{\mu}$ and which is orthogonal to {\it both} worldlines: $0 =
u_{\mu} b^{\mu} = \ell_{\mu} b^{\mu}$. This {\it bi-normal} $b^{\mu}$
is unique and is obtained by projecting $z^{\mu} (\xi) - y^{\mu}
(\tau)$ orthogonally to the two-plane spanned by $u^{\mu}$ and
$\ell^{\mu}$. Its origin $y^{\mu} (\tau_b)$ on the c.m. worldline
defines the (proper) time of impact: $t_0 = \tau_b$. By working in a
c.m. frame (with $u^{\mu} = (1,0,0,0)$ but $y_0^i$ not necessarily
zero) one easily finds
\begin{mathletters}
\label{eqn1}
\begin{eqnarray}
t_0 & = & \tau_b = z_0^0 - \ell^{-2} \ell^0 \ell^j (z_0^j - y_0^j) \, ,
\label{eqn1a} \\
b^0 & = & 0 \, , \label{eqn1b} \\
b^i & = & z_0^i - y_0^i - \ell^{-2} \ell^i \ell^j (z_0^j - y_0^j) \, ,
\label{eqn1c}
\end{eqnarray}
\end{mathletters}
which allows one to compute the derivatives with respect to
$z_0^{\lambda} = (z_0^0 , z_0^i)$ of $t_0$ and $b = \sqrt{\delta_{ij}
b^i b^j}$. [Note that $\partial t_0 / \partial z_0^i = -\ell^{-2}
\ell^0 \ell^i = -\ell^{-1} \ell^i$ does not vanish.] As for the
dependence on $z_0^i$ of the tensor projection
\begin{equation}
D_{11} (t_0) - D_{22} (t_0) = 2D_{11} + D_{33} - D_{ii} = 2 D_{ij}
\widehat{b}^i \widehat{b}^j + D_{ij} \widehat{\ell}^i
\widehat{\ell}^j - D_{ij} \delta^{ij} \, , \label{eqn2}
\end{equation}
where $\widehat{b}^i \equiv b^i / b$, $\widehat{\ell}^i \equiv \ell^i /
\ell$, it comes (besides the $z_0^i$-dependence of the time-argument
$t_0$) from the $b^i$-dependence of the term $2 D_{ij} \widehat{b}^i
\widehat{b}^j$ in Eq.~(\ref{eqn2}). By so computing the
$z_0^{\mu}$-derivative of Eq.~(\ref{eq3.33}), we verified the above
direct calculation of $\alpha_{\mu}$.

Equations (\ref{eq3.28}), (\ref{eq3.29}),
(\ref{eq3.30}) and (\ref{eq3.33}) are the main results of this paper.

\section{Discussion and conclusions}

Our explicit results (\ref{eq3.28})--(\ref{eq3.30}) show that the
suggestions Refs.~\cite{labeyrie,fakir,durrer} are
incorrect because the time-dependent part of the light deflection by a
localized gravitational source falls off as the {\it inverse cube} of
the impact parameter $b$, instead of their suggested $\alpha \sim h(b)
\propto b^{-1}$. Not only is the effect of the main $1/r$ retarded wave
cancelled, but even the subleading retarded contribution $\propto
1/r^2$ has no effect. This implies that the effect of the local
gravitational source will be much too small (for reasonable impact
parameters, when considering chance alignments) to be of observational
interest. We concentrated above on astrometric effects (geometrical
deflection), but our negative conclusions apply equally well to
photometric effects (scintillation) which can be directly derived from
the redshift and deflection effects we computed.

Note that something rather remarkable happened in our calculations.
Though we performed them for technical convenience in Fourier space,
the quantity we evaluated is the line integral (\ref{eq2.3}) in which
$h_{\mu \nu} (x^{\lambda})$ is the full (quadrupolar) retarded field
given by Eqs.~(\ref{eq3.21}), with the structure $h(t,r) \sim a_1 (t-r)
/ r + a_2 (t-r) / r^2 + a_3 (t-r) / r^3$. The final results
(\ref{eq3.28})--(\ref{eq3.30}) not only depend on the fastest decaying
contribution $a_3 (t-r) / r^3$, but they no longer contain an integral
over time. Without our making a near-zone approximation (in which one
expands all the retarded quantities, $a(t-r/c) = a(t) - \frac{r}{c}
\dot a (t) + \frac{r^2}{2c^2} \ddot a (t) + \cdots$) the exact results
depend only on the value of the coefficient $a_3$ at the time $t_0$ of
closest impact. [In particular, as we said above the results do not
depend on the retarded, advanced or time-symmetric Green function
used.]

In other words, our results can be stated by saying that the exact
deflection in the complicated retarded field is simply obtained by
computing the deflection in the $t_0$-instantaneous near-zone
gravitational field
\begin{equation}
h_{00}^{\rm near \ zone} (t,{\bf x}) = \left[ \frac{2GM}{r} \right] +
G \partial_{ij} \frac{D_{ij} (t)}{r} + \cdots \, , \label{eq4.34}
\end{equation}
\begin{equation}
h_{ij}^{\rm near \ zone} (t,{\bf x}) = h_{00}^{\rm near \ zone}
\delta_{ij} \, . \label{eq4.35}
\end{equation}
Here, we have added (in brackets) to the previously considered
time-dependent quadrupolar part, the static monopolar part associated
to the total mass $M$ of the gravitational source, which causes a
well-known deflection $\alpha_1 = -4 GM/b$. Taking advantage of this
dependence on the $t_0$-instantaneous near-zone field, it is possible
to reexpress our results (\ref{eq3.28})-(\ref{eq3.33}) in a very
compact (but somewhat subtle) way by considering the scalar potential
(\ref{eq3.31}) due to a unit-mass monopolar field, $h_{\mu\nu}^1 =
2G\delta_{\mu\nu}/r$. One finds
\begin{equation}
V^{\rm mono}_1( b ) = -4G\ln b \, , \label{mono1}
\end{equation}
after discarding a formally infinite additional constant which
is irrelevant in view of the later application of derivatives to
$V^{\rm mono}_1(b)$.

As the two spatial derivatives (acting on $r^{-1} = \vert {\bf x} -
{\bf y}_0 \vert$) in the quadrupolar term in Eq.~(\ref{eq4.34}) can be
replaced by derivatives with respect to the c.m. $y_0^i$, we can very
simply express the total, monopolar plus quadrupolar, scalar potential
in terms of the unit-mass quadrupolar one
\begin{equation}
V^{\rm tot} (z_0^{\lambda}) = M V_1^{\rm mono} (b) +
\frac{1}{2} D_{ij} (t_0)
\frac{\partial^2}{\partial y_0^i \partial y_0^j}
V_1^{\rm mono} (b) \, .
\label{eqn3}
\end{equation}
It is easily checked (using Eqs.~(\ref{eqn1}) to differentiate $\ln b$)
that the quadrupolar piece of Eq.~(\ref{eqn3}) yields back
Eq.~(\ref{eq3.33}). Finally, using the general result (\ref{eq3.32}),
the deflection 4-vector $\alpha_{\mu}$ can be written entirely as a sum
of derivatives of the unit-mass monopolar potential (\ref{mono1}).

In view of our present, ``negative'' results (absence of large enough
time-dependent deflections) we did not study in as much detail the
effects of the higher multipole moments. We just formally checked [by
inserting in Eq.~(\ref{eq2.19}) the expansion in powers of ${\bf k}$
of $\widehat{T}_{ij} (\omega , {\bf k})$] that their contributions
fall off with $b$ at least as fast (and probably faster) than
the quadrupolar one.

Let us also note in passing that our results can be easily extended to
the case where gravity is not described by Einstein's theory but by a
more general tensor-scalar theory. Indeed, let us work in the
``Einstein conformal frame'' in which the field equations read (see,
e.g., \cite{def92})
\begin{equation}
R_{\mu \nu} = 2 \partial_{\mu} \varphi \partial_{\nu} \varphi + 8 \pi G
\left( T_{\mu \nu} - \frac{1}{2} g_{\mu \nu} T \right) \, ,
\label{eq4.36}
\end{equation}
\begin{equation}
\Box \varphi = - 4 \pi G \alpha (\varphi) T \, . \label{eq4.37}
\end{equation}
At linearized order in the deviations from a flat background
$\eta_{\mu \nu}$ with constant background value of the scalar field
$\varphi_0$, i.e. writing $g_{\mu \nu} = \eta_{\mu \nu} + h_{\mu
\nu}$, $\varphi = \varphi_0 + \phi$, the field equations become
simply (in harmonic gauge)
\begin{equation}
\Box h_{\mu \nu} = -16 \pi G (T_{\mu \nu} - \frac{1}{2} \eta_{\mu \nu}
T) \, ,
\label{eq4.38}
\end{equation}
\begin{equation}
\Box \phi = -4 \pi G \alpha (\varphi_0) T \, . \label{eq4.39}
\end{equation}
Moreover, null geodesics are conformally invariant so that light rays
are null geodesics in the Einstein metric $g_{\mu \nu}$ (as well as in
the Jordan-Fierz metric $\widetilde{g}_{\mu \nu} = A^2 (\varphi) g_{\mu
\nu}$, where the conformal factor $A(\varphi)$ is related to the
$\varphi-$dependent coupling $\alpha(\varphi)$ of Eq.~(\ref{eq4.37}) by
$\alpha(\varphi) = d \ln A(\varphi)/d\varphi$). The first-order
geometrical deflection of a light ray is then given by the same
Eq.~(\ref{eq2.3}) as above. As the first order tensor-scalar field
equation (\ref{eq4.38}) determining $h_{\mu \nu}$ is equivalent to the
first-order Einstein equation (\ref{eq2.8}) (and that $\partial_{\nu}
T^{\mu \nu} = 0$ holds also to that order of approximation) the total
deflection $\Delta \ell_{\mu}$ will be given by the same formula in
tensor-scalar theories as in Einstein theory. Differences would appear
only at the second order where the term $\partial_{\mu} \phi
\partial_{\nu} \phi$ starts contributing. Note that, strictly speaking,
the value and time-dependence of the quadrupole moment $D_{ij} (t)$ can
differ at lowest (``Keplerian'') order when considering, e.g., a binary
system made of neutron stars \cite{def92}. However, this does not
change the main conclusion that the deflection from localized
gravitational source falls off as $b^{-3}$. In our present framework
(where both the light source and the observer are ``at infinity'', i.e.
in a place where the deviations from the background $(\eta_{\mu \nu} ,
\varphi_0)$ are neglected), there is no (time-dependent) difference at
the observer between the Einstein metric, and the more ``physical''
Jordan-Fierz metric $\widetilde{g}_{\mu \nu}$. Therefore our result
$\Delta \ell_{\mu}^{\rm total} \propto b^{-3}$ in the Einstein
conformal frame implies the same conclusion for the physically measured
deflection $\Delta \widetilde{\ell}_{\mu}$. The situation would be
slightly more subtle if the deviations $h_{\mu \nu}$ and $\phi$ could
not be neglected at the location of the observer (or the light source).
See below our mention of edge effects in Einstein's theory. [Note that
when discussing photometric effects, if $\phi$ cannot be neglected near
the observer, one must take into account the additional area changes
and variable redshifts which enter when translating Einstein-frame
results into physical-frame ones.] Anyway, the main point of the
present work is to discuss the importance of localized gravitational
sources happening to lie close to the line of sight, and our framework
is sufficient to show that these locally generated effects are much
smaller than one might a priori think.

Let us also mention some useful consequences of our results. First,
one could think that there remains (barring very improbable exact
chance alignments) one class of physical systems where the light rays
would propagate through the near zone field of a gravitational
source, namely, that where the light source is located within the
gravitational source. For instance, we can think of a binary system,
of which one body is emitting electromagnetic radiation. [A binary
pulsar is precisely a system of this type.] Though our calculation
does not really apply to such a system, it suggests very strongly
that all the radiative pieces of the gravitational field (and, in
particular, the slowly decreasing $1/r$ emitted retarded wave) do not
contribute to the light deflection. The latter can be simply computed
by using a $t_0$-instantaneous, static approximation to the near zone
field. This confirms that the existing calculations of the local
gravitational time delay (the integral of $\alpha^0$) in binary
pulsars \cite{dd86}, which used such an approximation, are accurate.

Finally, another consequence of our calculations is that, in the real
case where neither the light source, nor the observer are ``at
infinity in a flat spacetime'', our results show that the observable
light deflection can be computed by neglecting localized gravitational
sources, and, more generally, any quasi-localized gravitational wave
packets (as we proved explicitly above, Eq.~(\ref{eq2.7n})). Therefore
the observable effect will come essentially from ``edge effects'',
i.e. from the fact that either the light source, or the observer
are actually embedded in a non localized background of gravitational
waves. This confirms the results of
Refs.~\cite{zipoy,zipoybertotti,bertottitrevese,kaiserjaffe}, and
shows that these effects can be correctly calculated (as was done in
these references) by neglecting all the source-rooted, near-zone-type
parts of the total gravitational field $h_{\mu \nu} (x)$ and
replacing $h_{\mu \nu} (x)$ by a pervading sea of on-shell, vacuum
wave packets.

\end{document}